\documentstyle[aps,twocolumn,epsfig]{revtex}

\begin{document}
\draft
\wideabs{
\title{Abelian threshold models and forced weakening}
\author{Eric F. Preston $^{\dag}$, Jorge S. S\'a Martins $^{\dag}$, and John B.
Rundle $^{\dag\star}$}
\address{$\dag$ Colorado Center for Chaos and Complexity, CIRES, CB 216, \\
University of Colorado, Boulder, CO 80309}
\address{$\star$ Geophysics Program, Physics Department, University of\\
Colorado, Boulder, CO 80309}
\date{\today}
\maketitle

\begin{abstract}
Mean field slider block models have provided an important entry point for
understanding the behavior of discrete driven threshold systems. We present
a method of constructing these models with an arbitrary frictional weakening
function. This `forced weakening' method unifies several existing
approaches, and multiplies the range of possible weakening laws. Forced
weakening also results in Abelian rupture propagation, so that an avalanche
size depends only on the initial stress distribution. We demonstrate how
this may be used to accurately predict the long-time event statistics of a
simulation.
\end{abstract}

} 

Complex spatial or temporal patterns often emerge when systems are driven
from equilibrium and experience instabilities. Coupled lattice maps have
become a popular method of modelling such phenomena, seeking to capture the
essential physics in a discrete form amenable to numerical simulation. In
the study of earthquakes, simple slider-block models \cite
{Burridge67,Rundle77,Nakanishi90,Ferguson97} have long been employed to
investigate the origin of magnitude-frequency scaling. Similar models are
used to describe such diverse phenomena as pinned charge density waves \cite
{Fisher85}, flux lattices in type II superconductors \cite{Giamarchi95}, and
creeping contact lines \cite{Ertas94}.

Recently, analytic results for the avalanche size distribution have been
presented for mean-field slider-block models \cite{Dahmen98,Ding93}. These
models are characterized by homogeneous, infinite range elastic
interactions, and exhibit complex event histories and regimes of behavior.
Despite their simplicity, mean-field models remain sensitive to the choice
of update rules and details of implementation. This is especially evident in
models that impose a particular form of simulated frictional weakening \cite
{Dahmen98}, where different modes of behavior appear as the strength of the
weakening is varied.

Here we describe a method of constructing a mean-field threshold model where
one may simulate arbitrary weakening laws under identical rules of
evolution. This unifies the analysis of previously incompatible models, and
provides more freedom in numerical simulation. This technique also results
in Abelian rupture propagation, where the size of a simulated earthquake is
uniquely determined from initial conditions, leading to a rigorous and
implementation-independent analysis.

The general slider-block model represents stick-slip motion along a fault
plane with $N\gg 1$ discrete coordinates (or `sites'). Each site $i$ is
assigned a slip deficit $u_{i}$, which measures the local distance from
elastic equilibrium. The sites are pinned in place by frictional forces, and
are subject to a restoring force (stress) proportional to their slip
deficit. Internal disorder gives rise to an additional component of stress
due to elastic interactions. The stress $\widetilde{\sigma }_{i}$ at a site $%
i$ is related to the slip deficits through a linear constitutive relation

\begin{equation}
\widetilde{\sigma }_{i}=-K_{L}u_{i}-\sum_{j}K_{ij}\left( u_{i}-u_{j}\right)
\end{equation}
where $K_{L}$ and $K_{ij}$ are spring constants. If we impose uniform
(mean-field) interactions between all the elements, $K_{ij}=K_{C}/N$, the
above relation becomes 
\begin{equation}
\widetilde{\sigma }_{i}=-K_{L}u_{i}-K_{C}(u_{i}-\left\langle u\right\rangle )
\label{cont eq with units}
\end{equation}
where $\ \left\langle u\right\rangle =N^{-1}\sum_{i}u_{i}$ will denote an
average over all $N$ sites in the model. We obtain a unitless expression by
dividing by $K_{C}a$, where $a$ is a characteristic microscopic length.
Defining the unitless slip deficit $\phi =u/a$, stress $\sigma =\widetilde{%
\sigma }/(K_{c}a)$, and spring constant ratio $K_{R}=K_{L}/K_{C}$, Eq. [\ref
{cont eq with units}] simplifies to 
\begin{equation}
\sigma _{i}=-(K_{R}+1)\phi _{i}+\left\langle \phi \right\rangle .
\label{const eq}
\end{equation}
For finite $N$ we will refer to this as the near mean-field (NMF) model.
Note that it is easy to invert Eq. (\ref{const eq}) for the slip deficits in
terms of stresses, 
\begin{equation}
\phi _{i}=\frac{-\sigma _{i}}{K_{R}+1}-\frac{\left\langle \sigma
\right\rangle }{K_{R}(K_{R}+1)}
\end{equation}
so the configuration is uniquely determined by the parameter $K_{R}$ and
either the slip deficits or stresses alone.

The model is slowly driven away from equilibrium by uniformly increasing the
slip deficits. Eventually the stress at one site will surpass the maximum
local frictional force and `fail', sliding toward its equilibrium point. The
motion of a failed site will change the mean slip deficit $\left\langle \phi
\right\rangle $, and produce a change in stress at other sites. If this
change brings other sites to their threshold, they will also fail, producing
an avalanche interpreted as a single event.

It is assumed that following the initiation of motion, the frictional force
on a site will weaken, producing a transient dynamic instability. In
discrete time we cannot model the dynamic slip or velocity of the site, but
instead assign a residual stress $\sigma ^{R}$ at which the motion arrests.
This $\sigma ^{R}$ is chosen from a probability distribution independently
for each failed site. Since slips occur instantaneously, we lose the
interplay between a continuously evolving stress field and frictional force
at a site. The behavior of dynamical models is known to strongly depend on
the form of frictional weakening \cite{Burridge67}, so the ability to
include equivalent effects in discrete models would be advantageous.

Since the NMF model is not dynamical we are only interested in large-scale
features of its behavior that are independent of microscopic dynamics. Thus
we are free to choose the simplest update rules that are consistent with the
phenomenon of interest. In practice, we assume that a single site $j$
reaches its stress threshold first. Since the physics will depend only on
changes in stress, we may impose a uniform failure threshold $\sigma ^{F}$
by absorbing any variations into the residual stress distribution. \ The
slip displacement $\Delta _{j}=\phi _{j}^{(f)}-\phi _{j}^{(i)}$ is related
to the change in stress $\Delta \sigma _{j}=\sigma _{j}^{R}-\sigma ^{F}$ by $%
\Delta _{j}=-\Delta \sigma _{j}/(K_{R}+1-N^{-1})$. The motion of the site
will change the mean slip deficit $\left\langle \phi \right\rangle $ by $%
\Delta _{j}/N$. We may view this as a transfer of stress from failing sites
to all others.

The above describes a series type dislocation where sites fail in sequence
and the stress transfer occurs to other sites all at once. This makes it
likely that any failing site (other than the single initiator) will have a
stress slightly {\em above} the threshold, which subtly provides an
order-of-failure dependence to the stress transfer. As a consequence, the
exact stress transfer in simulation will depend on obscure factors like the
order of iteration over sites. To eliminate this we must examine the stress
transfer in more detail.

Suppose that in the course of a (possibly ongoing) event there have been $k$
block failures. Let $\{k\}$ represent the set of indices of failed sites.
Call $\kappa =k/N$ the fraction of failed sites. Then from Eq. (\ref{const
eq}) the change in stress for any stable site $i$ is

\begin{eqnarray}
\Delta \sigma _{i\notin \{k\}} &=&\frac{1}{N}\sum_{j\in \{k\}}\Delta _{j}=%
\frac{\delta }{N}\sum_{j\in \{k\}}(\sigma _{j}^{f}-\sigma _{j}^{R}) 
\nonumber \\
&=&\delta \kappa \left( \left\langle \sigma ^{f}\right\rangle
_{k}-\left\langle \sigma ^{R}\right\rangle _{k}\right)
\label{Stress Transfer 1}
\end{eqnarray}
where $\delta =(K_{R}+1-1/N)^{-1}$, and $\left\langle \cdot \right\rangle
_{k}$ is an average applied over failed sites. We call this the {\em %
external stress transfer} to signify that it applies to sites that are not
part of the rupture. The quantity $\sigma _{j}^{f}$ is the stress of site $j$
{\em at failure}, which may be greater than $\sigma ^{F}$. Note that the
slip displacement $\Delta _{j}$ is only dependent on the stress drop at
failure because pinning occurs immediately, and subsequent stress changes
will not affect the slip. Since $\sigma ^{f}$ is typically very near $\sigma
^{F}$ and the $\sigma _{j}^{R}$ are identically distributed random
variables, the term in parenthesis is on average independent of $k$. Thus
the external transfer grows linearly (on average) with the fraction of
failed sites.

In `dynamic weakening' models \cite{Ben-Zion93} the increased propensity for
a failed site to slip further (due to a weakened pinning force) is simulated
by imposing a lower threshold stress $\sigma ^{D}<\sigma ^{F}$ for the
duration of a single event. After a site fails it will receive stress
transfer from subsequent failures, and thus may reach this lower threshold
and fail again. Re-failing sites will contribute more to the external
transfer and enhance the likelihood of continued rupture growth.

This form of weakening is strange in that it first requires some failed
sites to have their stress brought back up to the dynamical threshold, which
will occur at some minimum rupture size. Following the onset of dynamical
weakening the additional stress transfer typically results in a runaway
event which fails every block in the system. This feature can be exploited
to produce characteristic events which always occur once the minimum size is
reached.

One way to visualize the effects of weakening is to examine the average
stress of sites that have failed as a function of rupture size (Figure 1).
Without weakening, the average stress of failed sites (the {\em average
internal stress}) will itself grow linearly like the external transfer (with
half the slope). However, with dynamic weakening, all failed sites with
stress $\geq \sigma ^{D}$ will fail again, putting a ceiling on the average
internal stress.

We claim that a more realistic approach would be to have failed sites shed a
certain fraction of the stress they receive after failure. Implementing this
`fractional weakening' would involve re-computing the slip of all failed
sites with each new failure. There is an easier way to accomplish this if we
note that the desired effect is to lower the slope of the average internal
stress (AIS) function. There is a way to invert this relationship, such that
the average internal stress function is {\em given} and the requisite slips
and stress transfer computed as a result. In essence, in place of solving
dynamical equations involving slip or velocity dependent friction, we can
impose the {\em effects} of the weakening as they appear in a discrete time
context. We call this approach `forced weakening'.

To perform this inversion we first observe the change in stress of a site $j$
as it depends on $k$, the number of failed sites. Let $\Delta _{j}^{k}$
denote the slip displacement of site $j$ after $k$ failures. If it is
nonzero it includes all block motion, including initial failure, additional
failures from weakening, or continuous sliding. Consider the system of
equations for the stress changes of the failed sites $j\in \{k\}$%
\begin{eqnarray}
\Delta \sigma _{j\in \{k\}}^{k} &=&\sigma _{j}^{k}-\sigma _{j}^{0}=-\delta
\Delta _{j}^{k}+\frac{1}{N}\sum_{i\neq j\in \{k\}}\Delta _{i}^{k}  \nonumber
\\
&=&-(K_{R}+1)\Delta _{j}^{k}+\frac{1}{N}\sum_{i\in \{k\}}\Delta _{i}^{k}.
\label{Local Drops}
\end{eqnarray}
The last line demonstrates the simple linear form of the relationship
between stress drops and slip displacements. This may be obtained via a
matrix with diagonal elements $N^{-1}-(K_{R}+1)$ and off-diagonal elements $%
N^{-1}$. This matrix may easily be inverted to obtain the slips $\Delta
_{j}^{k}$ in terms of the current stress drops $\Delta \sigma _{j}^{k}$%
\begin{eqnarray}
\Delta _{j}^{k} &=&\frac{-(K_{R}+1-k/N)\Delta \sigma
_{j}^{k}-N^{-1}\sum_{i\in \{k\}}\Delta \sigma _{i}^{k}}{(K_{R}+1)\left(
K_{R}+1-k/N\right) }  \nonumber \\
&=&\frac{-\Delta \sigma _{j}^{k}}{K_{R}+1}-\frac{(k/N)\ \left\langle \Delta
\sigma ^{k}\right\rangle _{k}}{(K_{R}+1)\left( K_{R}+1-k/N\right) }
\label{Stress Transfer 2}
\end{eqnarray}
This expression provides the slips $\Delta _{j}^{k}$ that are necessary to
generate a set of given stress drops.

The slips are directly related to the external transfer, as in Eq. \ref
{Stress Transfer 1}. Summing over them yields a new expression for the
external transfer in terms of the stress drops. 
\begin{eqnarray}
\Delta \sigma _{i\notin \{k\}} &=&\frac{1}{N}\sum_{j\in \{k\}}\Delta _{j}=%
\frac{k}{N}\left\langle -\Delta \sigma ^{k}\right\rangle _{k}  \nonumber \\
&&\times \left[ \frac{1}{K_{R}+1}+\frac{k/N}{(K_{R}+1)(K_{R}+1-k/N)}\right] 
\nonumber \\
&=&\frac{\kappa }{(K_{R}+1-\kappa )}\left\langle -\Delta \sigma
^{k}\right\rangle _{\kappa }  \label{final stress transfer}
\end{eqnarray}
This is identical to Eq. (\ref{Stress Transfer 1}) when $k=1$. However, the
effective transfer coefficient $\delta =\delta (\kappa )=(K_{R}+1-\kappa
)^{-1}$ now grows with the rupture size. Making up for this is the fact that
the stress drops are no longer frozen after failure and will now decrease as
the rupture grows. Observe that to calculate the external transfer, we need
not specify individual stress drops or slips, but we just need to know the 
{\em average dynamic stress drop} of all failed sites for a given rupture
size. This evolving average stress drop is defined as 
\begin{eqnarray}
\left\langle -\Delta \sigma ^{k}\right\rangle _{\kappa } &=&\frac{1}{k}%
\sum_{j\in \{k\}}(\sigma _{j}^{0}-\sigma _{j}^{k})  \nonumber \\
&=&\left\langle \sigma ^{0}\right\rangle _{\kappa }-f(\kappa )
\label{final stress drop}
\end{eqnarray}
where we have defined the average internal stress (AIS) function $f(\kappa
)=\left\langle \sigma ^{k}\right\rangle _{k}$. The stress transfer now
depends only on the initial stress and the AIS function, which are
independent of failure order. Using this formulation, avalanches are Abelian.

In numerical simulation, we must eventually assign an actual slip and/or
residual stress to each failed site. Care must be taken to make results
agree with the analogous forward simulation. For example, given a linear AIS
function $f(\kappa )=\alpha \kappa +\beta $, we could assign the $k^{th}$
failed site a random residual stress (with mean $\beta $) plus $2\alpha
(k-1)/N$. When using AIS functions with no forward equivalent, the method of
assigning final stresses must be stated explicitly.

The forced weakening method has two main advantages. Practically, it allows
the simulation of models with arbitrary weakening characteristics, some of
which would not be obtainable with modified CA rules. Formally, the model is
Abelian, so that the event size is a unique function of the initial stress
configuration. Using this fact we can seek an expression which will
determine the event size given adequate information of the initial stresses.

To accomplish this we first approximate the stress configuration with a
continuous distribution $p_{\sigma }(\chi )$, where $p_{\sigma }(\chi )d\chi 
$ is the probability that a randomly selected site will have stress $\sigma $
between $\chi $, and $\chi +d\chi $. There are two ways of obtaining a
continuous probability distribution from a discrete set of stresses: allow
the number of sites $N$ to become infinite, or to consider averages over a
course grained time. In the latter case the continuous driving of the slip
deficits will quickly fill in the spectrum of possible stresses.

When considering a course grained time, we realize that the distribution is
only an average, and that the actual instantaneous distribution of stresses
will fluctuate. Similarly, the solution we obtain will be an average with
statistical properties we would like to calculate. Under ordinary
conditions, the distribution of stresses is statistically stationary \cite
{Preston01}, so the expected distribution of event sizes at any time will
predict the long-time average behavior.

To write down the next-event-size expression, it is convenient to define the
stress deficit $\Sigma =\sigma ^{F}-\sigma $, and the cumulative
distribution 
\begin{equation}
P_{\Sigma }(\chi )=\int_{0}^{\chi }p_{\Sigma }(\chi ^{\prime })d\chi
^{\prime }
\end{equation}
where $p_{\Sigma }(\chi )=p_{\sigma }(\sigma ^{F}-\chi )$. Given this
distribution the event size $\kappa $ is the solution to 
\begin{eqnarray}
\frac{1}{K_{R}+1-\kappa }\left[ \kappa \sigma ^{F}-\kappa f(\kappa
)-\sigma ^{F}\int_{0}^{\kappa }P_{\Sigma }(\chi )d\chi \right]  \nonumber \\
-\sigma ^{F}P_{\Sigma }^{-1}(\kappa +d\kappa ) &=&0  \label{stress excess}
\end{eqnarray}
where $f(\kappa )$ is again the AIS function. A full derivation and analysis
of the results of this equation will be published elsewhere. For an example
solution, suppose each stress in the system is chosen independently from a
uniform distribution between zero and $\sigma ^{F}$. Then for large $N$, Eq.
(\ref{stress excess}) determines that $\kappa $ is the first crossing of a
1-d random walk constrained to return to zero at $\kappa =1$. Without that
constraint, this produces a power law distribution of event sizes with
exponent $-3/2$. With the constraint in place, there is an upturn at large
event sizes, which is observed in simulations (so long as $K_{R}$ is low
enough to allow such large events). A comparison of this analytical solution
with a simulation is shown in Figure 2.

In summary, the forced weakening method extends the basic slider block model
to include arbitrary weakening functions in an efficient discrete time
simulation. Additionally, forced weakening results in Abelian avalanches
which leads to a rigorous and implementation-independent analysis. We have
presented a brief example of this showing excellent agreement between theory
and simulation. This approach should be equally applicable to related
discrete threshold models. While the resulting formalism is dependent on the
mean-field character of the model, the numerical techniques may find wider
use whenever an inversion like Eq. (\ref{Stress Transfer 2}) is available.

The authors thank W. Klein for his input. E.F.P. and J.B.R. acknowledge
support by DOE grant DE-FG03-95ER14499; J.S.S.M. was supported as a Visiting
Fellow by CIRES, University of Colorado at Boulder.

\begin{figure}[tbp]
\begin{center}
\includegraphics[angle=0,scale=0.6]{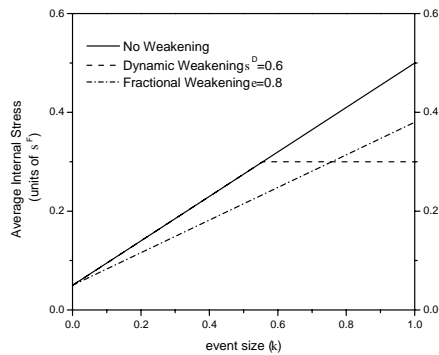}
\end{center}
\caption{Average internal stress (AIS) functions for several weakening laws.
The AIS function measures the average stress of failed sites as a function
of rupture size. `Dynamic weakening' as implemented in the literature places
an upper bound on the AIS starting at the dynamic threshold $\protect\sigma %
^{D}$. Fractional weakening, proposed here, sheds a fixed fraction of the
internal transfer to unfailed sites. In this figure we have taken the
average residual stress to be 0.05.}
\label{AIS funcs}
\end{figure}

\begin{figure}[tbp]
\begin{center}
\includegraphics[angle=0,scale=1.1]{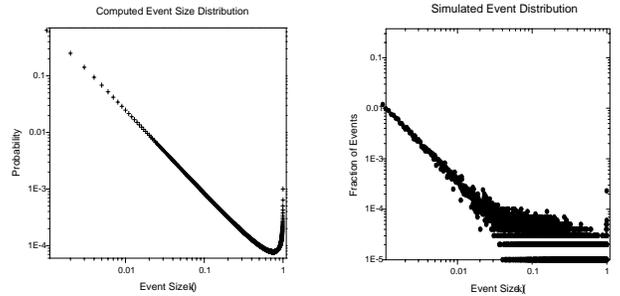}
\end{center}
\caption{Comparison of a solution of Eq. (\ref{stress excess}) for uniformly 
distributed stresses (a) with simulation results (b). 
The power laws have identical mean-field exponents of -3/2 and
the predicted upturn at large event sizes is evident in the simulated data.
The parameters of the simulation are $K _{R}=0.01$ and an average residual 
stress of 0.05. }
\label{comparison}
\end{figure}


\begin{references}
\bibitem{Burridge67}  R. Burridge and L. Knopoff, Bull. Seismol. Soc. Am. 
{\bf 57}, 341 (1967)

\bibitem{Rundle77}  J. B. Rundle and D. D. Jackson, Bull. Seismol. Soc. Am. 
{\bf 67}, 1363 (1977)

\bibitem{Nakanishi90}  H. Nakanishi, Phys. Rev. A {\bf 43}, 6613 (1990)

\bibitem{Ferguson97}  C. Ferguson, Ph.D. thesis, Boston University

\bibitem{Fisher85}  D. S. Fisher, Phys. Rev. B {\bf 31}, 1396 (1985)

\bibitem{Giamarchi95}  T. Giamarchi and P. Le Doussal, Phys. Rev. B {\bf 52},
1242 (1995)

\bibitem{Ertas94}  D. Ertas and M. Kardar, Phys. Rev. E {\bf 49}, R2532
(1994)

\bibitem{Dahmen98}  K. Dahmen, D. Ertas, and Y. Ben-Zion, Phys. Rev. E {\bf %
58}, 1494 (1998)

\bibitem{Ding93}  E. J. Ding and Y. N. Lu, Phys. Rev. Let. {\bf 70}, 3627

\bibitem{Ben-Zion93}  Y. Ben-Zion and J. R. Rice, J. Geophys. Res. {\bf 98},
14,109 (1993)

\bibitem{Preston01}  E. F. Preston, J. S. S\'{a} Martins, and J. B. Rundle,
in preparation
\end{references}
\end{document}